\documentclass[useAMS,usenatbib]{mn2e}

\usepackage[psamsfonts]{amssymb}
\usepackage[dvips]{graphicx}
\usepackage{graphicx}
\usepackage{amsmath,alltt}                                                                                                                          
\usepackage{multirow}
\usepackage{rotating}
\usepackage{lscape}    
\usepackage{tablefootnote}   
\usepackage{hyperref} 
\usepackage[english]{babel}                                                                                                                  

\title[Radial Variation in Stellar Mass Functions]{Radial Variation in the Stellar Mass Functions of Star Clusters}
\author[Webb $\&$ Vesperini]{Jeremy J. Webb \& Enrico Vesperini
\thanks{E-mail: jerjwebb@iu.edu (JW), evesperi@indiana.edu (EV)} \\
Department of Astronomy, Indiana University, Swain West, 727 E. 3rd Street, IN 47405 Bloomington, USA}

\begin{document}
\newcommand{\dalph}{\delta\alpha}
\newcommand{\galph}{\alpha_G}

\pagerange{\pageref{firstpage}--\pageref{lastpage}} \pubyear{2016}

\maketitle

\label{firstpage}

\begin{abstract}

A number of recent observational studies of Galactic globular clusters have measured the variation in the slope of a cluster's stellar mass function $\alpha$ with clustercentric distance $r$. In order to gather a deeper understanding of the information contained in  such observations, we have explored the evolution of $\alpha(r)$ for star clusters with a variety of initial conditions using a large suite of $N$-body simulations. We have specifically studied how the time evolution of $\alpha(r)$ is affected by initial size, mass, binary fraction, primordial mass segregation, black hole retention, an external tidal field, and the initial mass function itself. Previous studies have shown that the evolution of $\galph$ is closely related to the amount of mass loss suffered by a cluster. Hence for each simulation we have also followed the evolution of the slope of the cluster's global stellar mass function, $\alpha_G$, and have shown that clusters follow a well-defined track in the $\alpha_G$-$d\alpha(r)/d(ln(r/r_m))$ plane.  The location of a cluster on the $\galph-d\alpha(r)/d(ln(r/r_m))$ plane can therefore constrain its dynamical history and, in particular, constrain possible variations in the stellar initial mass function. The $\alpha_G$-$d\alpha(r)/d(ln(r/r_m))$ plane thus serves as a key tool for fully exploiting the information contained in wide field studies of cluster stellar mass functions.

\end{abstract}


\section{Introduction} \label{intro}

The two mechanisms which dominate the dynamical evolution of any globular cluster are stellar evolution and two-body relaxation. For globular clusters which survive their formation environment \citep[e.g.][]{gieles06, kruijssen11,kruijssen12,renaud13, rieder13}, stellar evolution is the initial driver behind cluster evolution and early expansion due to mass loss from high-mass stars. After this early stage, two-body relaxation takes over and the cluster evolves towards a state of partial energy equipartition \citep{merritt81, miocchi06, trenti13, gieles15, bianchini16}. A natural result of two-body relaxation is the segregation of higher mass stars towards the cluster center while lower mass stars migrate outwards \citep[e.g.][]{heggie03}. Two-body relaxation will also eventually lead to cluster dissolution as stars are energized to velocities greater than the cluster's escape velocity and pushed beyond the tidal radius $r_t$ imposed by the tidal field of the host galaxy. In the presence of a non-static external tidal field, tidal heating and shocks serve to inject energy and further accelerate the dissolution process \citep[e.g.][]{weinberg94, vesperini97, gnedin99, baumgardt03, webb13,webb14a,webb14b,brockamp14}. 

At early times, evolution of the stellar mass function is only affected by stellar evolution as massive stars lose mass. The early loss of outer stars associated with the expansion triggered by stellar evolution mass loss will not affect the mass function unless the cluster is primordially mass segregated \citep[e.g.][]{baumgardt08, vesperini09, haghi15}. However as a cluster undergoes mass segregation, the mean mass of escaping stars will decrease over time which will cause the slope of the mass function to flatten (become less negative). The combined effects of stellar evolution, two-body relaxation, and the presence of an external tidal field on the stellar mass function have been well studied throughout the literature \citep[e.g.][]{vesperini97,kruijssen09b, baumgardt03, trenti10, leigh12, lamers13,webb14a, webb15}. These studies have shown that the \textit{global} stellar mass function is primarily a function of the fraction of mass lost by a cluster, with some scatter about the relation due to the amount of mass segregation a cluster undergoes before it starts to lose a significant amount of stars. Additional factors that have been shown to at least partially influence the evolution of the stellar mass function as a function of fraction of mass lost include the total dissolution time of the cluster \citep{kruijssen09b}, the initial binary fraction and initial mass function (IMF) \citep{webb15}, the retention fraction of dark remnants  \citep{trenti10, lutzgendorf13}, and the presence of an intermediate mass black hole \citep{lutzgendorf13}. 

In \citet{webb14a}, the effects of an external tidal field on the evolution of the stellar mass function \textit{at different clustercentric radii} was explored for clusters with circular and eccentric orbits. The slope of the mass function expectedly increases in the inner regions and decreases in the outer regions as stars within the cluster segregate. \citet{webb14a} also found that an external field will influence the mass function beyond the half-mass radius $r_m$ by preferentially stripping stars in the outer regions which, as a result of the cluster's evolution towards partial energy equipartition, are dominated by low-mass stars. 

With continuing advancements in telescope capabilities and observational techniques, it has become possible to reliably measure the mass function of Galactic globular clusters at different clustercentric radii. Such measurements allow for a study of the radial variation of the mass function, and have been done recently for M10 \citep{beccari10}, Pal 4 \citep{frank12}, Pal 14 \citep{frank14}, NGC 6101 \citep{dalessandro15}, NGC 5466 \citep{beccari15}, and 47 Tuc \citep{zhang15}. The general picture outlined in \citet{webb14a} is observed in M10, Pal4, Pal 14, NGC 5466 and 47 Tuc as all five clusters show clear signs of mass segregation (although to different degrees). For four of the clusters, the degree to which they are mass segregated is commensurate with their respective half-mass relaxation times $t_{rh}$ and the effects of tidal stripping. Pal 14 on the other hand, despite a high present day $t_{rh}$, also shows clear evidence of mass segregation in its $\alpha$ profile \citep{frank14}. The authors therefore suggest that Pal 14 must have either been partially mass segregated at birth or was much more compact in the past. Curiously, the $\alpha$ profile in NGC 6101 is nearly constant suggesting that cluster has undergone no mass segregation. The observed differences between M10, Pal 4, Pal 14, NGC 6101, NGC 5466 and 47 Tuc suggest that additional studies on how the radial variation of the stellar mass function evolves with time, and its dependence on various cluster parameters, are required before the $\alpha$ profile of a cluster can be used to constrain its formation conditions and dynamical history. 

In this study, we present the evolution of the radial variation in the stellar mass function for a large suite of $N$-body model clusters with different initial conditions. In Section \ref{s_nbody} we discuss the full suite of simulations used in this study. In Section \ref{s_mfms} we first explore how variations in $\alpha$ with clustercentric distance r ($\alpha(r)$) evolve as a function of time and fraction of mass lost for two specific model clusters in isolation. Then we identify the effects of an external tidal field, initial size, the IMF, primordial mass segregation, primordial binary fraction, initial cluster mass, orbit and black hole retention on the evolution of $\alpha(r)$. We also consider whether the mass range over which the slope of the stellar mass function is measured introduces any type of bias. Varying initial cluster mass, size, and orbit further allows us to explore how the evolution of $\alpha(r)$ depends on a cluster's tidal filling factor $\frac{r_m}{r_t}$. Each of our findings are discussed and summarized in Section \ref{s_discussion}, where we also identify three distinct stages in the evolution of $\alpha(r)$.

\begin{table*}
  \caption{Model Input Parameters}
  \label{table:modparam}
  \begin{center}
    \begin{tabular}{lccccccc}
      \hline\hline
      {Name} & {Mass} & {$r_{m,i}$} & {$R_p$} & {e} & {S} & {$\frac{N_{B,i}}{N_{tot}}$} & {IMF} \\
      \hline

Isolated Clusters \\
\hline
IRM1 & $3\times10^4 M_\odot$ & 1.1 pc & n/a & n/a & 0 & 0 & K93 \\
IRM6 & $3\times10^4 M_\odot$ & 6 pc & n/a & n/a & 0 & 0 & K93 \\

\hline
Clusters in External Tidal Fields \\
\hline
Standard Tidally Filling and Under-filling Models\\
\hline
RM1 & $6\times10^4 M_\odot$ & 1.1 pc & 6 kpc & 0 & 0 & 0 & K93 \\
RM6 (IMFK93, NB0) & $6\times10^4 M_\odot$ & 6 pc & 6 kpc & 0 & 0 & 0 & K93 \\

\hline
Primordial Mass Segregation\\
\hline

PMS0/IMFK01 & $6\times10^4 M_\odot$ & 6 pc & 6 kpc & 0 & 0 & 0 & K01 \\
PMS10 & $6\times10^4 M_\odot$ & 6 pc & 6 kpc & 0 & 0.1 & 0 & K01 \\
PMS25 & $6\times10^4 M_\odot$ & 6 pc & 6 kpc & 0 & 0.25 & 0 & K01 \\
PMS50 & $6\times10^4 M_\odot$ & 6 pc & 6 kpc & 0 & 0.5 & 0 & K01 \\

\hline
IMF\\
\hline

IMF185 & $6\times10^4 M_\odot$ &  6 pc & 6 kpc & 0 & 0 & 0 & $\alpha_0 = 1.85$ \\
IMF235 & $6\times10^4 M_\odot$ &  6 pc & 6 kpc & 0 & 0 & 0 & $\alpha_0 = 2.35$ \\
IMFBPL & $6\times10^4 M_\odot$ &  6 pc & 6 kpc & 0 & 0 & 0 & $\alpha_0=-0.9$ for $0.1 < m < 0.5 M_\odot$ \\
{}&{}&{}&{}&{}&{}&{}& $\alpha_0=-2.3$ for $0.5 \le m \le 50  M_\odot$. \\

\hline
Binary Fraction\\
\hline

NB2 & $6\times10^4 M_\odot$ & 6 pc & 6 kpc & 0 & 0 & $2\%$ & K93 \\
NB4 (BH0, E0R6RM6) & $6\times10^4 M_\odot$ & 6 pc & 6 kpc & 0 & $4\%$ &0 & K93 \\

\hline
Mass \tablefootnote{\citet{webb13}, \citet{leigh13}, \citet{webb15}} \\
\hline
M30K & $3\times10^4 M_\odot$ & 6 pc & 6 kpc & 0 & 0 & $4\%$ & K93 \\
M60K & $6\times10^4 M_\odot$ & 6 pc & 6 kpc & 0 & 0 & $4\%$ & K93 \\
M80K & $8\times10^4 M_\odot$ & 6 pc & 6 kpc & 0 & 0 & $4\%$ & K93 \\
M110K & $1.1\times10^5 M_\odot$ & 6 pc & 6 kpc & 0 & 0 & $4\%$ & K93 \\

\hline
Size and Orbit \tablefootnote{\citet{leigh13}, \citet{webb15}} \\
\hline

E0R6RM1 & $6\times10^4 M_\odot$ & 1.1 pc & 6 kpc & 0 & 0 & $4\%$ & K93 \\
E05RP6RM1 & $6\times10^4 M_\odot$ & 1.1 pc & 6 kpc & 0.5 & 0 & $4\%$ & K93 \\
E0R18RM1 & $6\times10^4 M_\odot$ & 1.1 pc & 18 kpc & 0 & 0 & $4\%$ & K93 \\
E05RP6RM6 & $6\times10^4 M_\odot$ & 6 pc & 6 kpc & 0.5 & 0 & $4\%$ & K93 \\
E0R18RM6 & $6\times10^4 M_\odot$ & 6 pc & 18 kpc & 0 & 0 & $4\%$ & K93 \\
E09RP6RM6 & $6\times10^4 M_\odot$ & 6 pc & 6 kpc & 0.9 & 0 & $4\%$ & K93 \\
E0R104RM6 & $6\times10^4 M_\odot$ & 6 pc & 104 kpc & 0 & 0 & $4\%$ & K93 \\

\hline
Black Hole Retention\\
\hline

BH25 ($\frac{N_{BH,retained}}{N_{BH,created}}=0.25$) & $6\times10^4 M_\odot$ & 6 pc & 6 kpc & 0 & 0 & $4\%$ & K93 \\
BH50 ($\frac{N_{BH,retained}}{N_{BH,created}}=0.5$) & $6\times10^4 M_\odot$ & 6 pc & 6 kpc & 0 & 0 & $4\%$ & K93 \\

\hline\hline

\multicolumn{2}{l}{$^1$ \citet{webb13}, \citet{leigh13}, \citet{webb15}} \\ 
\multicolumn{2}{l}{$^2$ \citet{leigh13}, \citet{webb15}} \\

    \end{tabular}
  \end{center}
\end{table*}

\section{N-body models} \label{s_nbody}

The 12 Gyr evolution of each model star cluster in our suite of simulations was simulated using the direct $N$-body code NBODY6 \citep{aarseth03}. While some of the models used here have been presented in previous studies, we will summarize again the general parameters of the entire suite of simulations for clarity and provide additional details for any new models. The initial radial profile of each cluster is generated from a Plummer density profile \citep{plummer11} out to a cut-off radius of 10 times the initial half-mass radius $r_{m,i}$ of the cluster. While we consider model clusters with a range IMFs (see Table \ref{table:modparam}), in all cases the minimum and maximum initial stellar masses are equal to 0.1 and 50 $M_\odot$ respectively and stellar metalicities are set equal to 0.001. For clusters with a non-zero initial binary fraction, total binary masses are set equal to the sum of two stars drawn from the IMF and the masses of individual stars within the binary are randomly selected from a uniform distribution. The stellar evolution algorithms for single and binary stars are discussed in \citet{hurley00} and \citet{hurley02} and the initial distribution of binary periods and orbital eccentricities are taken from \citet{duquennoy91} and \citet{heggie75} respectively. Clusters are modelled either in isolation or orbiting within a Milky Way-like potential made up of a $1.5 \times 10^{10} M_{\odot}$ point-mass bulge, a $5 \times 10^{10} M_{\odot}$ \citet{miyamoto75} disk (with $a=4.5\,$ kpc and $b=0.5\,$ kpc), and a logarithmic halo potential \citep{xue08} that is scaled to ensure that the circular velocity at a galactocentric distance $R_{gc}$ of $8.5\,$ kpc is 220 km/s.   

The first two model clusters for which we study the evolution of $\alpha(r)$ have initial masses of $3\times10^4 M_\odot$, $r_{m,i}$ of 1.1 pc and 6 pc and no primordial binaries. The initial distribution of stellar masses follows a \citet{kroupa93} IMF (K93), with

\begin{equation}\label{eqn:mfunc}
\frac{dN}{dm} = m^\alpha
\end{equation}

and $\alpha$ equals -2.7 for $m > 1 M_\odot$, -2.2 for $0.5 \le m \le 1  M_\odot$, and -1.3 for $0.08 < m \le 0.5  M_\odot$. We initially model these two clusters in isolation (IRM1 and IMR6) to set a basis for comparing how different initial conditions alter the evolution of $\alpha(r)$. To study the effects of an external tidal field, we then model each cluster with a circular orbit at 6 kpc in the Milky Way-like potential. Due to the higher escape rates experienced by clusters in an external tidal field, all the subsequent model clusters have a larger initial number of particles and  masses of $6\times10^4 M_\odot$ (unless otherwise specified). When orbiting at 6 kpc in the Milky Way-like potential, the $r_{m,i} = 1.1$ pc cluster (RM1) is considered to be our standard tidally under-filling cluster with $\frac{r_m}{r_t} = 0.04$, while the $r_{m,i} = 6$ pc cluster (RM6) is considered to be our standard tidally filling cluster with $\frac{r_m}{r_t} =0.2$.  

To explore how other initial parameters may affect the evolution of $\alpha(r)$, we have run simulations of the standard tidally filling and under-filling cases but with a range of IMFs, initial degrees of mass segregation (S), initial binary fractions ($\frac{N_{B,i}}{N_{tot}}$), masses, orbits, and black hole retention fractions ($\frac{N_{BH,retained}}{N_{BH,created}}$). Table \ref{table:modparam} contains a complete list of all the models used in this study. To generate model clusters with different IMFs, we have used the publicly available code McLuster \citep{kupper11} to simulate clusters with power law IMFs ($\alpha_0 =1.85$,  $\alpha_0 = 2.35$), a \citet{kroupa01} IMF (K01), and a broken power law (BPL) IMF ($\alpha_0=-0.9$ for $0.1 < m < 0.5 M_\odot$ and $\alpha_0=-2.3$ for $0.5 \le m \le 50  M_\odot$). McLuster was also used to generate primordially mass segregated clusters, which were assumed to have a \citet{kroupa01} IMF so they could be compared to the aforementioned K01 model. McLuster imposes mass segregation following the method introduced in \citet{baumgardt08}, which ensures that model clusters with different degrees of primordial mass segregation have the same density profile. Black hole retention is incorporated into our models by ensuring that the desired retention fraction of black holes are not given a velocity kick upon creation.   

\section{Evolution of the Radial Variation in the Stellar Mass Function}\label{s_mfms}
 
To measure the radial variation of the stellar mass function at each time step for each model, special care was taken to properly bin the data with respect to both clustercentric distance and mass. \citet{maiz05} found that a variable bin size, which ensures that the same number of stars in each mass bin, minimizes any numerical bias when determining a cluster's mass function. We also apply this approach to our division of the cluster into radial bins.

 \begin{figure*}
\centering
\includegraphics[width=\textwidth, height =16 cm]{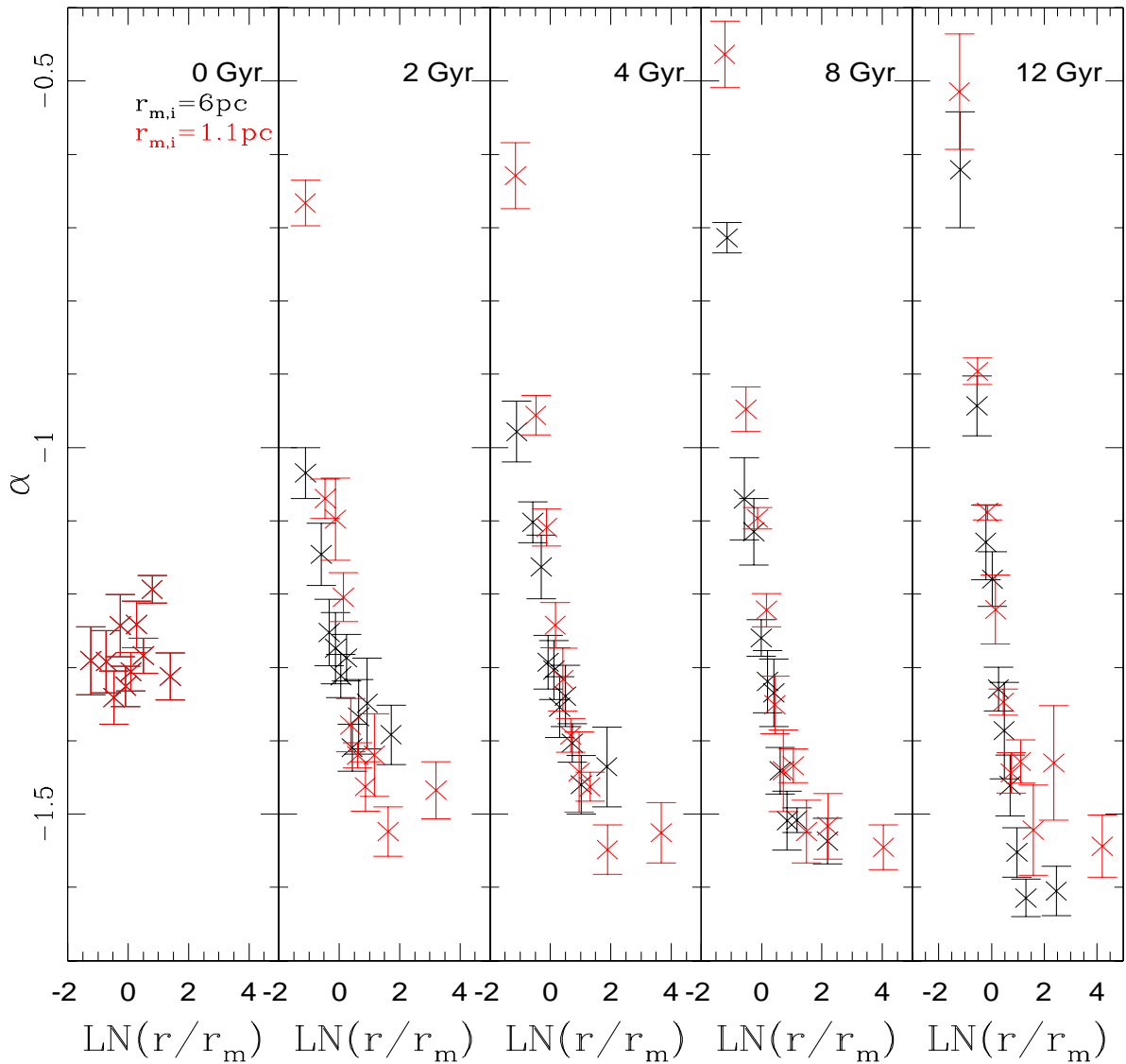}
\caption{Slope of the stellar mass function for stars between 0.1- 0.5 $M_\odot$ at different clustercentric radii (normalized by $r_m$) for isolated globular clusters with initial masses of $3 \times 10^4 M_\odot$ and initial half mass radii of 1.1 pc (red) and 6 pc (black).}
  \label{fig:aprof}
\end{figure*}

We initially calculate the global stellar mass function for all stars within 0.1 and 0.5 $M_\odot$. Stars are sub-divided into 6 different mass bins, with the bin sizes (dm) kept variable in order to have the same number of stars dN in each mass bin. The mean mass of all stars in a given mass bin is used to set the corresponding value of m. The slope of the global stellar mass function $\alpha_G$ is then set equal to the slope of a plot of $\log(\frac{dN}{dm})$ versus $\log(m)$. Next, we sub-divide all stars with masses between 0.1 and 0.5 $M_\odot$ into 10 radial bins such that each bin contains $10\%$ of the sub-sample of stars. The mean clustercentric radius of all the stars in a given radial bin is used to set the corresponding value of r. $\alpha$ is then calculated within each radial bin, yielding $\alpha(r)$ at each time step.

\subsection{Star Clusters in Isolation}\label{sec:isolated}

Simulations of clusters in isolation yield the evolution of $\alpha(r)$ when stellar evolution and two-body relaxation are the only mechanisms driving cluster evolution. For our standard isolated clusters IRM1 and IRM6, we plot the value of $\alpha(r)$ between $0.1- 0.5$ $M_\odot$ as a function of the natural logarithm of each radial bin at 0, 2, 4, 8 , and 12 Gyr in Figure \ref{fig:aprof}. Each radial bin has been normalized by $r_m$ to better compare clusters of different size. 

Figure \ref{fig:aprof} illustrates the effect that mass segregation
has on the slope of the mass function at different clustercentric
distances. As massive stars migrate inwards and low-mass stars migrate
outwards, $\alpha(r)$ will flatten in the inner regions and steepen
(become more negative) in the outer regions of the cluster. Since
$\alpha(r)$ is also affected by the radial variation of the local
relaxation time, which increases with clustercentric distance, the
inner $\alpha(r)$ profile will evolve more rapidly than the outer
one. This second point is why the $\alpha(r)$ profile is not a linear
function of clustercentric distance. 

\begin{figure}
\centering
\includegraphics[width=\columnwidth]{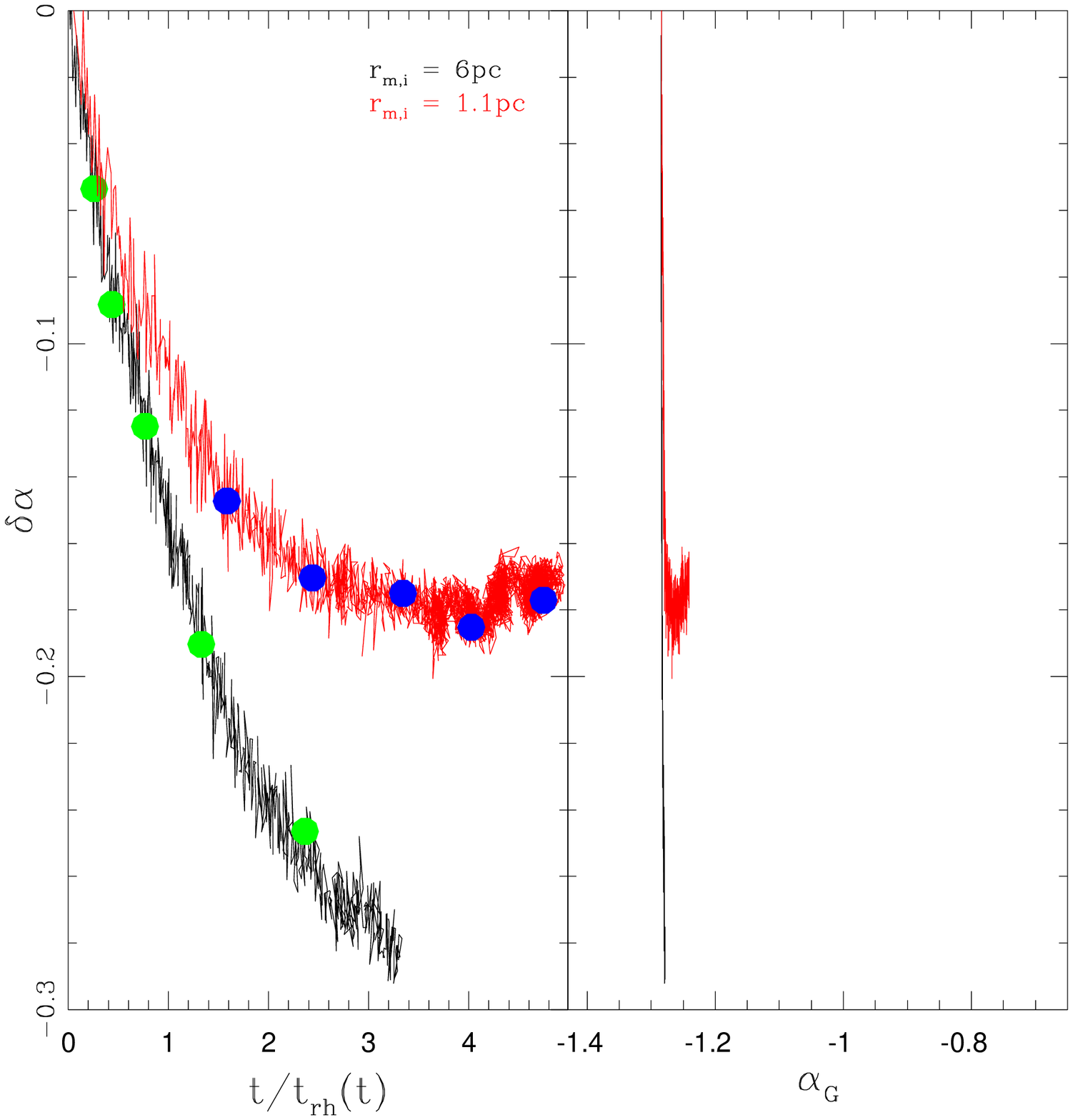}
\caption{Slope of the radial variation in the stellar mass function for stars between 0.1- 0.5 $M_\odot$ as a function of time normalized by current relaxation time (left panel) and the global $\alpha_G$ (right panel) for isolated globular clusters with initial masses of $6.3 \times 10^4 M_\odot$ and initial half mass radii of 1.1 pc (red) and 6 pc (black). The blue and green filled circles mark 0.5 Gyr, 1 Gyr, 2 Gyr, 4 Gyr and 8 Gyr of evolution for the 1.1 pc and 6 pc model clusters respectively.}
  \label{fig:dalphai}
\end{figure}

To quantify the evolution of radial variations in the stellar mass
function, we calculate the
slope,$\frac{d\alpha(r)}{d(ln(\frac{r}{r_m}))}$, of the line of best
fit to $\alpha(r)$ versus $\ln (\frac{r}{r_m})$ at each time step;
hereafter we refer to this slope simply as $\dalph$. Figure
\ref{fig:dalphai} illustrates the evolution of $\dalph$ as a function
of both time (normalized by the cluster's instantaneous $t_{rh}(t)$)
and $\galph$. The normalization of time by $t_{rh}(t)$, calculated
using the formalism of \citet{spitzer71}, serves as a proxy
  for comparing clusters at similar dynamical ages as can be
  determined by observers.

As discussed in the Introduction, a number of studies have shown that
$\alpha_G$ traces the fraction of stars lost by a cluster; hence
comparing the evolution of $\dalph$ to $\galph$ allows us to establish
a connection between two fundamental aspects of a cluster's dynamical
evolution: internal mass segregation and stellar escape. In
the specific case of our simulations of isolated clusters,
escape is negligible and the only effect shown in this figure
is that of internal mass segregation. 

The left panel of Figure \ref{fig:dalphai} shows the time evolution of
$\dalph$ and demonstrates how mass segregation expectedly leads to a
progressive increase in the radial variation of $\alpha$. Initially,
as a cluster expands and undergoes mass segregation $\dalph$ will
decrease. The rate at which $\dalph$ decreases initially
  scales with a cluster's $t_{rh}$, however differences in the
  structural evolution of the compact and extended clusters leads to
  the divergence of the two evolutionary tracks.
 In particular the smaller $t_{rh}$ of the compact system
  implies that when it reaches the same value of
  $\frac{t}{t_{rh}(t)}$ as the extended system it has an earlier age and therefore a
  broader range of stellar masses; as a consequence of this, stars
  with masses between 0.1 and 0.5 $M_{\odot}$ are less segregated in
  the compact system than in the extended one. To illustrate this
point we show in the left panel of Figure 1 the points corresponding
to a few different ages on the tracks of the two systems.

 Once core-collapse occurs, mass segregation stops while the
relaxation time at all clustercentric distances is increasing,
consistent with the findings of a number of previous studies showing
that mass segregation either stops or significantly slows down in the
post-core collapse phase \citep[e.g.][]{giersz96}. During this second
phase $\dalph$ remains approximately constant since mass segregation
has stopped, with the possibility of a mild increase due to the
post-core collapse expansion. Since the extended cluster has not reached 
core collapse by the end of the simulation and $\dalph$ is still decreasing, the different
levels of mass segregation (as measured by $\dalph$) reached by the
two clusters after 12 Gyr is a consequence of their different
dynamical histories and core-collapse times. The rate at which
$\dalph$ decreases for the extended cluster does however slow done,
due the isolated cluster continuing to expand, so $\dalph$ will not
continue to decrease indefinitely. As shown in the right panel of
Figure \ref{fig:dalphai}, even though these clusters are initially
losing mass due to the stellar evolution of high-mass stars the
global mass function \text{for stars} between 0.1 and 0.5
$M_\odot$ remains unaffected over 12 Gyr since they evolve in
isolation and very few stars actually escape the cluster.  

While the results of this section serve to illustrate the fundamental aspects of cluster internal dynamical evolution, clusters do not actually evolve in isolation and will lose stars over time. Hence we must turn to the more realistic case of globular clusters orbiting in a gravitational field in order to determine how $\dalph$ evolves as a function of $\galph$ when both relaxation and tidal stripping are affecting cluster evolution.

\subsection{Star Clusters Evolving in an External Tidal Field}\label{sec:tidal}

\begin{figure}
\centering
\includegraphics[width=\columnwidth]{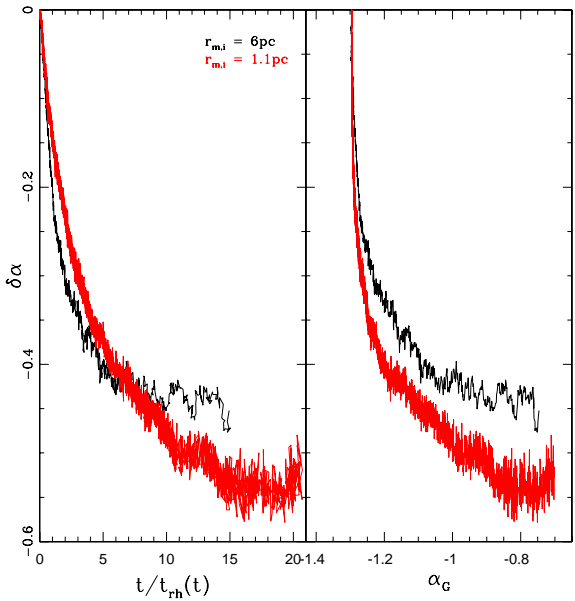}
\caption{Slope of the radial variation in the stellar mass function for stars between 0.1- 0.5 $M_\odot$ as a function of time normalized by current relaxation time (left panel) and $\alpha_G$ (right panel) for globular clusters with circular orbits at 6 kpc, initial masses of $6.3 \times 10^4 M_\odot$, and initial half mass radii of 1.1 pc (red) and 6 pc (black).}
  \label{fig:dalpha}
\end{figure}

In order to explore how the presence of an external tidal field will alter the evolution of $\dalph$, we take the $r_{m,i} =$ 1.1 and 6 pc clusters that were previously evolved in isolation and study their evolution assuming they move on circular orbits at 6 kpc in a Milky Way-like tidal field. The evolution of $\dalph$ with respect to both $\frac{t}{t_{rh}(t)}$ and $\alpha_G$ is illustrated in Figure \ref{fig:dalpha}.

An external tidal field has the primary effect of limiting the maximum
size each cluster can reach and significantly enhance the rate of
stellar escape compared to the isolated case. The resulting change to
the evolution of $\dalph$ is two-fold. Since each cluster will be
smaller than if it were in isolation, the $t_{rh}$ of each cluster is
shorter and $\dalph$ will decrease at a faster rate. Later, as
$\dalph$ decreases and the mean mass of escaping stars begins to
decrease, the preferential escape of low-mass stars slows the decrease
of $\alpha$ in the outer regions which in turn slows the decrease in
$\dalph$. The preferential escape of low-mass stars also causes
a progressive flattening in the global mass function, as
clearly illustrated by the evolution of $\dalph$ with respect to
$\galph$ in the right panel Figure \ref{fig:dalpha}. Eventually the
influence of low-mass stars being stripped from the cluster on
$\dalph$ becomes greater than the influence of low-mass stars
segregating outwards, such that $\alpha(r)$ in the outer regions of
the clusters starts to increase. Once $\alpha(r)$ begins increasing in
the outer regions, $\dalph$ will have reached its minimum value and
will remain constant until the cluster reaches dissolution.  

\subsection{The Stellar Mass Function Over Different Mass Ranges and Timescales}

\begin{figure}
\centering
\includegraphics[width=\columnwidth]{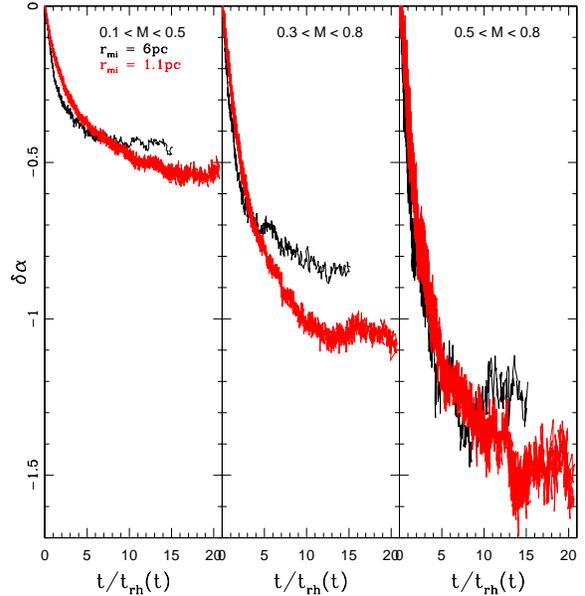}
\caption{Slope of the radial variation in the stellar mass function for stars between 0.1 and 0.5 $M_\odot$ (left panel),  0.3 and 0.8 $M_\odot$ (center panel) and 0.5 and 0.8 $M_\odot$ (right panel) as a function of dynamical time for globular clusters with circular orbits at 6 kpc, initial masses of $6 \times 10^4 M_\odot$, and initial half mass radii of 1.1 pc (red) and 6 pc (black).}
  \label{fig:dalpha_mranget}
\end{figure}

\begin{figure}
\centering
\includegraphics[width=\columnwidth]{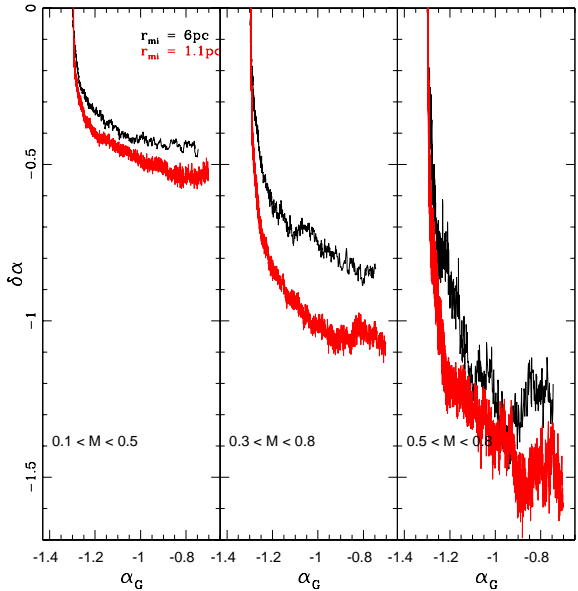}
\caption{Slope of the radial variation in the stellar mass function for stars between 0.1 and 0.5 $M_\odot$ (left panel),  0.3 and 0.8 $M_\odot$ (center panel) and  0.5 and 0.8 $M_\odot$ (right panel) as a function of $\alpha_G$ for globular clusters with circular orbits at 6 kpc, initial masses of $6 \times 10^4 M_\odot$, and initial half mass radii of 1.1 pc (red) and 6 pc (black).}
  \label{fig:dalpha_mrangea}
\end{figure}

We next consider whether the mass range over which the slope of the stellar mass function is measured will influence the evolution of $\dalph$ as outlined in Sections \ref{sec:isolated} and \ref{sec:tidal}. In Figures \ref{fig:dalpha_mranget} and \ref{fig:dalpha_mrangea} we show the evolution of $\dalph$ for the standard tidally filling and under-filling model clusters but with $\alpha$ measured over two other commonly used mass ranges: 0.3 - 0.8 $M_\odot$ and 0.5 - 0.8 $M_\odot$.   

From Figures \ref{fig:dalpha_mranget} and \ref{fig:dalpha_mrangea} we see that, as expected, stars in the higher mass ranges segregate more rapidly and reach a higher level of mass segregation than stars in lower mass ranges. Hence when measured using stars with $0.5 < m < 0.8 M_\odot$, $\dalph$ is able to reach a lower minimum value before segregation slows compared to when stars between $0.1$ and 0.5 $M_\odot$ are used. However, the evolution of $\dalph$ is still qualitatively the same regardless of the mass range used to measure $\alpha(r)$.

\subsection{The Initial Mass Function}
\begin{figure}
\centering
\includegraphics[width=\columnwidth]{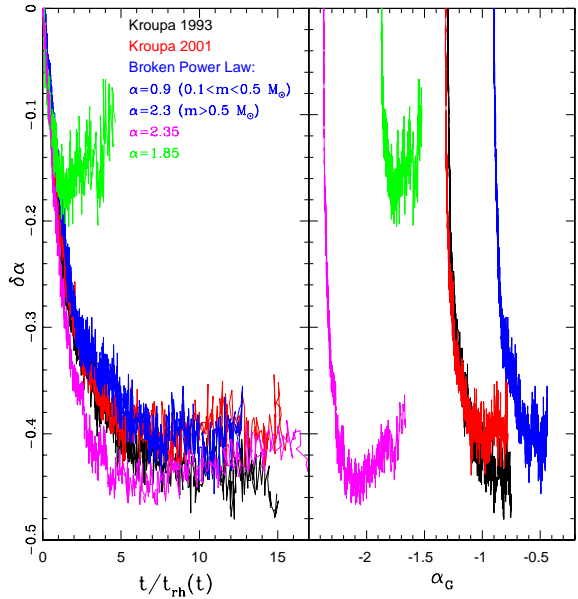}
\caption{Slope of the radial variation in the stellar mass function for stars between 0.1- 0.5 $M_\odot$ as a function of time normalized by current relaxation time (left panel) and the global $\alpha_G$ (right panel) for globular clusters with circular orbits at 6 kpc, initial masses of $6.3 \times 10^4 M_\odot$, initial half mass radii 6 pc and different functional forms of their IMF.}
  \label{fig:imf}
\end{figure}

In this section we study how the functional form of the IMF affects the evolution of
$\alpha(r)$. We make this comparison in Figure \ref{fig:imf}, where we
consider a Kroupa (2001) IMF, two power-law IMFs and a broken power-law IMF. 
 
Figure \ref{fig:imf} illustrates that dramatically changing the IMF of
a cluster can alter the evolution of $\dalph$. More specifically, if a
different IMF results in a significant change to the cluster's
$t_{rh}$ or its structural evolution then the evolution of $\dalph$ will change accordingly. Using the
extreme case of a power-law IMF with $\alpha_{G,0} = -1.85$ as an
example (IMF185), the large population of high-mass stars results in
mass loss due to stellar evolution causing the cluster to undergo
significant expansion at early times. A larger number of stars are therefore lost during the expansion process compared to the K93 case, with the cluster reaching a tidal filling factor greater than 0.325 when expansion due to stellar evolution is complete. Hence IMF185 will experience a higher stellar escape rate than the more massive K93 model \citep{gieles11}, and $\alpha_G$ will start evolving at a higher (less negative) value of $\delta_\alpha$. However it should be noted that a tidal filling factor greater than 0.325 is larger than all of the Galactic clusters except Pal 5 \citep{baumgardt10}.

For clusters that have IMFs that are more similar to IMFK93, such that there is not a major difference in their structural evolution, the evolution of $\dalph$ with respect to $\frac{t}{t_{rh}(t)}$ appears to be independent of the functional form of the IMF.  Model clusters IMFK01 ($\alpha_0=-1.3$ for $0.1 < m < 0.5 M_\odot$ and $\alpha_0=-2.3$ for $0.5 \le m \le 50  M_\odot$), IMF235 ($\alpha_{G,0} = -2.35$) and IMFBPL ($\alpha_0=-0.9$ for $0.1 < m < 0.5 M_\odot$ and $\alpha_0=-2.3$ for $0.5 \le m \le 50  M_\odot$) all segregate at similar rates as IMFK93. An observable difference between these model clusters can instead be found in the evolution of $\dalph$ with respect to $\galph$. Clusters with different IMFs start with different initial values of $\galph$ such that they are initially offset in the right panel of Figure \ref{fig:imf}. Since the relationship between $\dalph$ and the amount that $\galph$ changes from its initial value remains independent of the IMF, model clusters with bottom heavy (IMF235) or top heavy (IMFBPL) IMFs will never simultaneous reach the same $\dalph$ and $\galph$ as K93 or K01. Hence the $\dalph$-$\galph$ plane can be used to disentangle variations in the mass function slope due to a non-universal IMF from variations due to dynamical evolution and the preferential escape of low-mass stars. 

\subsection{Primordial Mass Segregation} 

\begin{figure}
\centering
\includegraphics[width=\columnwidth]{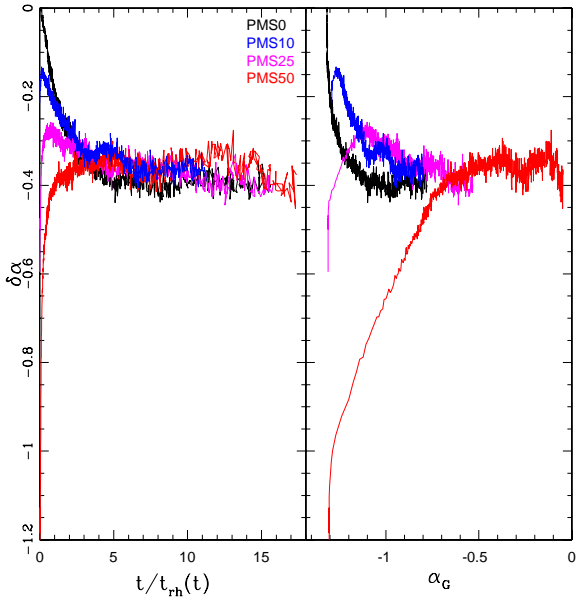}
\caption{Slope of the radial variation in the stellar mass function for stars between 0.1 and 0.5 $M_\odot$ as a function of time normalized by current relaxation time (left panel) and the global $\alpha_G$ (right panel) for globular clusters with circular orbits at 6 kpc, initial masses of $6.3 \times 10^4 M_\odot$, initial half mass radii 6 pc and primordial segregation factors of 0 (black), 0.25 (magenta), 0.5 (red) and 0.9 (red).}
  \label{fig:mseg}
\end{figure}

A key assumption made in our previous models is that globular clusters are not primordially mass segregated and have an initial $\dalph$ equal to 0. Not only will the initial $\dalph$ of primordially mass segregated be non-zero, but they will undergo a larger initial expansion and therefore lose more stars via tidal stripping than a cluster with no primordial mass segregation \citep[e.g.][]{baumgardt08, vesperini09, haghi15}. To explore the effects of primordial mass segregation on the evolution $\dalph$, we have modelled the evolution of our $r_{m,i} =$ 6 cluster with a K01 IMF while adding varying degrees of primordial mass segregation (Figure \ref{fig:mseg}).

Figure \ref{fig:mseg} illustrates that, as expected, a primordially mass segregated cluster will start its evolution with a more negative $\dalph$ and is generally characterized by a more negative $\dalph$ in the early phases of its evolution compared to a non-segregated cluster. Since primordially mass segregated clusters undergo a significant early expansion associated with mass loss due to stellar evolution \citep[e.g.][]{baumgardt08, vesperini09, haghi15}, $\dalph$ will initially increase. When this expansion occurs, a large number of low-mass stars escape the cluster which also causes $\galph$ to increase rapidly. It should be noted however that the global mass function over higher mass ranges ( $0.3 < m < 0.8 M_\odot$ or $0.5 < m < 0.8 M_\odot$) will evolve more slowly since it is mainly low-mass stars that initially escape the clusters.

After the initial expansion phase, $\dalph$ begins to decrease for the moderately segregated clusters (PMS10, PMS25) as each cluster resumes the mass segregation process. For the highly segregated cluster (PMS50), the early expansion results in a $t_{rh}$ so long that the cluster is unable to segregate any further after its initial expansion.

The fact that each of these simulations converge to the same value of $\dalph$ in the left panel of Figure \ref{fig:mseg} suggests that any effect primordial mass segregation has on the evolution of $\dalph$ is lost after clusters begin losing stars via tidal stripping. However, the right panel of Figure \ref{fig:mseg} suggests the $\dalph$-$\galph$ plane could be used to identify clusters with stronger primordial mass segregation during the early stages of their evolution. Before $\dalph$ reaches the asymptotic value of the unsegregated model, primordially mass segregated clusters will appear to have undergone significant mass segregation while maintaining a relatively unchanged mass function.

\subsection{Initial Binary Fraction} 

\begin{figure}
\centering
\includegraphics[width=\columnwidth]{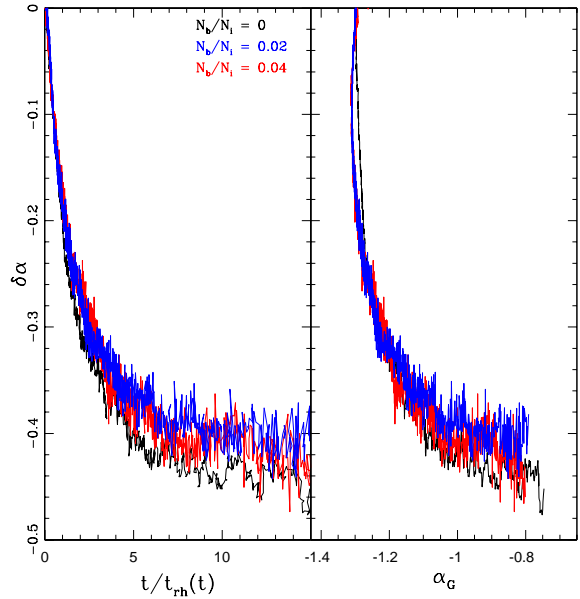}
\caption{Slope of the radial variation in the stellar mass function for stars between 0.1- 0.5 $M_\odot$ as a function of time normalized by current relaxation time (left panel) and the global $\alpha_G$ (right panel) for globular clusters with circular orbits at 6 kpc, initial masses of $6.3 \times 10^4 M_\odot$, initial half mass radii 6 pc and initial binary fractions of $4\%$ (black), $2\%$ (blue) and $0\%$ (red).}
  \label{fig:binary}
\end{figure}

Since globular clusters form with non-zero initial binary fractions, it is very important to determine whether the presence of primordial binaries can alter the evolution of $\dalph$. To address this issue, we have modelled the evolution of the tidally filling cluster (NB0) with initial binary fractions of $2\%$ (NB2) and $4\%$ (NB4). Figure \ref{fig:binary} shows that the time evolution of $\dalph$ is not significantly affected by the presence of primordial binaries. Since binaries minimally affect neither the expansion rate or stellar escape rate of a cluster (and therefore do not affect its $t_{rh}$), $\dalph$ for two clusters that have the same age can be directly compared without having to worry about whether or not the clusters had different binary fractions at birth.

Binaries do however influence the early measurement of $\alpha_G$, as
discussed in \citet{webb15}. By treating binaries as unresolved,
binary stars will initially appear as single high mass stars. As
binary stars become unbound, an apparent influx of additional low-mass
stars will occur causing $\alpha_G$ to initially decrease before
increasing as the cluster loses mass. This is however a small effect and becomes
negligible after clusters have lost $50\%$ of their initial mass (see
Webb \& Leigh 2015).
 
\subsection{Initial Cluster Mass}\label{s_masses}

\begin{figure}
\centering
\includegraphics[width=\columnwidth]{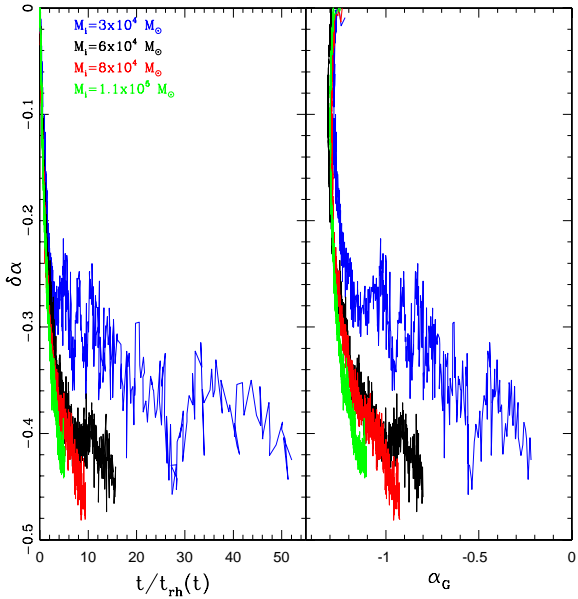}
\caption{Slope of the radial variation in the stellar mass function for stars between 0.1- 0.5 $M_\odot$ as a function of time normalized by current relaxation time (left panel) and the global $\alpha_G$ (right panel) for globular clusters with circular orbits at 6 kpc, initial masses between $3 \times 10^4 M_\odot$ and $1.1 \times 10^5 M_\odot$, and initial half mass radii of 6 pc. Different model clusters are colour coded based on mass, as indicated by the legend.}
  \label{fig:dalpha_mtot}
\end{figure}

With an understanding that primordial binaries minimally affect the evolution of $\dalph$, we can make use of previous simulations with initial binary fractions of $4\%$ to study how initial cluster mass can influence the evolution of $\dalph$. Initially presented in \citet{webb13}, \citet{leigh13} and \citet{webb15}, we have additional simulations of the tidally filling case with $4\%$ binaries (M60K) but with initial masses between $3 \times 10^4 M_\odot$ and $1.1 \times 10^5 M_\odot$ (M30K, M80K, and M110K). The evolution of $\dalph$ for each of these models is presented in Figure \ref{fig:dalpha_mtot}. 

Since these models have the same $r_{m,i}$ and are evolved at a fixed galactocentric radius, models with different masses have different tidal filling factors, relaxation times, and star-escape/lifetime timescales. Differences in the evolution of $\delta_\alpha$ versus $\frac{t}{t_{rh}(t)}$ (see the left panel of Figure 9) are therefore a consequence of the different rate of the structural and mass evolution for clusters with different initial masses. The higher mass clusters are more under-filling and lose mass at a slower rate than the low-mass clusters, such that they have more time to segregate before escaping stars cause $\alpha(r)$ to stop decreasing in the outer regions (in agreement with Section \ref{sec:tidal}). Lower mass clusters on the other hand are more filling and characterized by smaller initial relaxation times, evolving towards even smaller values of $t_{rh}(t)$ after the initial expansion stage. Hence low-mass clusters reach a given value of $\frac{t}{t_{rh}(t)}$ earlier than higher mass clusters and are characterized by a smaller degree of mass segregation. Moreover, with the higher mass loss rate also experienced by lower mass clusters, $\alpha(r)$ stops decreasing in the outer regions of the cluster (causing the evolution of $\delta_\alpha$ to slow down as well) at a lower $\frac{t}{t_{rh}(t)}$ compared to the high mass clusters. The relative widths of each clusters mass function do not play a significant role in the evolution of each $\delta_\alpha$ here, as it did in Section \ref{sec:isolated}, since the evolutionary track of the lowest mass cluster is identical to the other models for the first 4 Gyr.

Differences in relaxation times and different star-escape/lifetime timescales also dictate the extent of the model evolution of $\dalph$ with respect to $\galph$. The more rapid rate of star escape in lower mass systems leads to a more rapid evolution towards flatter slopes for the mass function (see the right panel of Figure \ref{fig:dalpha_mtot}). Hence $\galph$ in systems with lower masses is able to start evolving before significant mass segregation has occurred (as measured by $\dalph$).

\subsection{Initial Size and Cluster Orbit}

\begin{figure}
\centering
\includegraphics[width=\columnwidth]{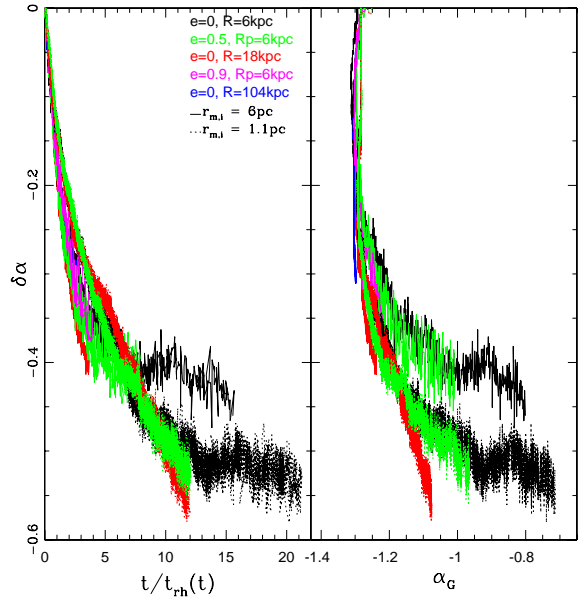}
\caption{Slope of the radial variation in the stellar mass function for stars between 0.1- 0.5 $M_\odot$ as a function of time normalized by current relaxation time (left panel) and $\alpha_G$ (right panel) for globular clusters with circular orbits between 6 and 18 kpc, orbital eccentricities between 0 and 0.9, initial masses of $6 \times 10^4 M_\odot$, and initial half mass radii of 1.1 pc (dotted lines) and 6 pc (solid lines). Different model clusters are colour coded based on orbit, as indicated by the legend.}
  \label{fig:dalpha_orbit}
\end{figure}

To explore how a cluster's tidal filling factor affects $\dalph$, we again make use of models from \citet{leigh13} and \citet{webb15} with initial binary fractions of $4\%$ and $r_{m,i} = 1.1$ pc and 6 pc. To compare with the standard tidal filling case (E0RP6RM6), we determine the evolution of $\dalph$ for identical model clusters with circular orbits at 18 kpc (E0RP18RM6) and 104 kpc (E0RP104RM6) and with eccentric orbits that have perigalactic distances of 6 kpc and orbital eccentricities (e) of 0.5 ((E05RP6RM6) and 0.9 (E09RP6RM6). To compare with the standard tidally under-filling case (E0RP6RM1), we follow the evolution of $\dalph$ for one model cluster with a circular orbit at 18 kpc (E0RP18RM1) and another with a perigalactic distance of 6 kpc and an orbital eccentricity of 0.5 (E05RP6RM1). The evolution of $\dalph$ with respect to $t/t_{\rm rh}(t)$ and $\galph$ for all 8 models is illustrated in Figure \ref{fig:dalpha_orbit}.

The interpretation of the evolutionary tracks shown in Figure \ref{fig:dalpha_orbit} follows our discussion of models with different initial masses in Section \ref{s_masses}. The evolution of $\dalph$ and $\galph$ is dictated by the internal relaxation time and the timescale of star loss. For example, two clusters on the same orbit (e.g. the model with $e=0$ and $R=6$ kpc) but different initial half-mass radii will lose stars at similar rates (dictated by the strength of the tidal field (see e.g. \citet{gieles08}), but the timescale of segregation will be shorter for the initially more compact cluster. As the two clusters lose stars and evolve towards smaller values of $\galph$, the more compact cluster will do
so while converging to a smaller value of $\dalph$ (i.e. a larger degree of mass segregation).

\subsection{Black Hole Retention} 

\begin{figure}
\centering
\includegraphics[width=\columnwidth]{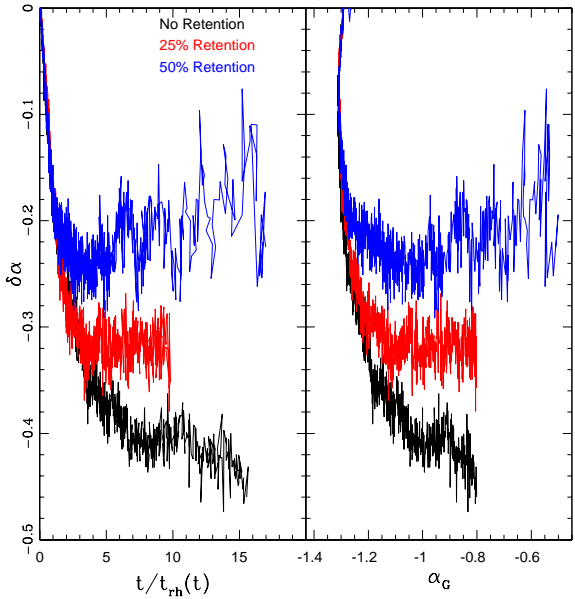}
\caption{Slope of the radial variation in the stellar mass function for stars between 0.1- 0.5 $M_\odot$ as a function of time normalized by current relaxation time (left panel) and the global $\alpha_G$ (right panel) for globular clusters with circular orbits at 6 kpc, initial masses of $6.3 \times 10^4 M_\odot$, initial half mass radii 6 pc and dark remnant retention fractions of 0 (black), 0.25 (blue) and 0.5 (red).}
  \label{fig:bh}
\end{figure}

\citet{trenti10} and \citet{lutzgendorf13} found that a population of  dark remnants (black holes, neutron stars) or a massive black hole at the centre of the cluster will increase the escape rate of high-mass stars that have migrated to the centre of the cluster via mass segregation. Hence the \textit{global} mass function will evolve at a slower rate as low-mass stars that have migrated outwards via mass segregation are not the only stars escaping the cluster. Furthermore, intermediate mass black holes have also been shown to quench mass segregation in globular clusters and may also influence the evolution $\dalph$ \citep{gill08, pasquato09}. To determine whether the evolution of $\dalph$ is affected by dark remnants, we have performed additional simulations with non-zero black hole retention fractions. More specifically we have re-simulated the tidally filling case with a circular orbit at 6 kpc and a binary fraction of $4\%$ (BH0), but with $25\%$ (BH25) and $50\%$ (BH50) of newly created black holes not given velocity kicks at their time of formation such that they are retained by the cluster. The evolution of $\dalph$ for all three models is illustrated in Figure \ref{fig:bh}.

From Figure \ref{fig:bh} we see that black hole retention can slow and even completely halt the evolution of $\delta_\alpha$, in agreement with previous studies \citep{trenti10, lutzgendorf13}. The retention of black holes affects the mass segregation process in two different ways. First, a sub-population of near stellar mass black holes will migrate inwards and cause the core of the host star cluster to be much larger than if no black holes were retained. Hence the cluster's core relaxation time will increase and $\alpha(r)$ will increase in the inner regions of the cluster at a slower rate. $\delta_\alpha$ will therefore be lower at a given $\frac{t}{t_{rh}(t)}$ for clusters with large black hole retention fractions.

Second, a sub-population of black holes will cause a cluster to initially expand to a larger $r_m$ than if no black holes were retained such that the cluster becomes more tidally filling \citep{merritt04, mackey08}. Expanding to a larger filling factor means more stars escape the cluster during the expansion phase and $\alpha(r)$ will begin evolving from its primordial value at an earlier age, effectively halting the mass segregation process. Therefore, clusters with higher black hole retention fractions stop segregating at higher values of $\delta_\alpha$ (less negative) than clusters that retain no black holes.

It should be noted that even if a cluster is able to retain a certain fraction of black holes when they first form, and have those black holes affect the evolution of its structural properties (and therefore $\delta_\alpha$), it does not mean the cluster contains each and every one of the retained black holes today. Dynamical evolution results in black holes continuously being ejected from the cluster's core. In fact, after 12 Gyr only $11\%$ and $17\%$ of all the black holes that were created remain in the model clusters with $25\%$ and $50\%$ retention fractions respectively. Hence a cluster will be more strongly affected by the retention of black holes than the present day black hole population would imply.

\subsection{Projection Effects and Application to the Milky Way} 

\begin{figure}
\centering
\includegraphics[width=\columnwidth]{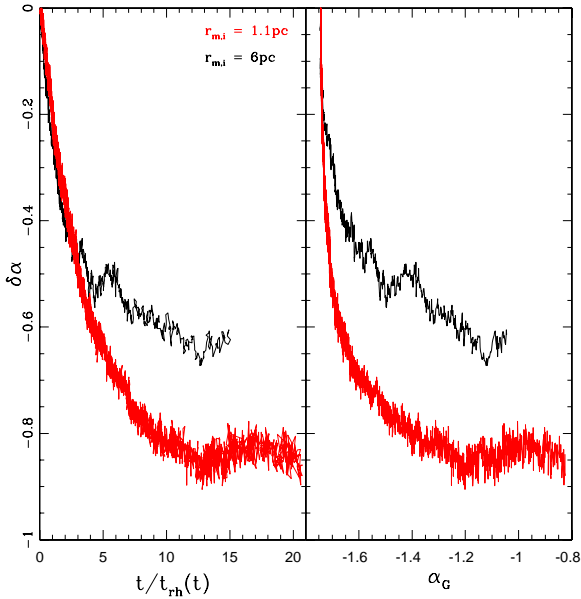}
\caption{Slope of the projected radial variation in the stellar mass
  function for stars between 0.3- 0.8 $M_\odot$ as a function of time normalized by current relaxation
  time (left panel) and $\alpha_G$ (right panel) for globular clusters
  with circular orbits at 6 kpc, initial masses of $6.3 \times 10^4
  M_\odot$, and initial half mass radii of 1.1 pc (red) and 6 pc
  (black).} 
  \label{fig:dalphap}
\end{figure}

In order to apply our models to observations of Galactic globular clusters, we must consider the evolution of $\delta_\alpha$ when measured in projection. We first take the $r_{m,i} =$ 1.1 and 6 pc model clusters on circular orbits at 6 kpc in a Milky Way-like tidal field (shown in Figure \ref{fig:dalpha}) and radially bin stars based on their projected clustercentric distance R. The line of best fit to $\alpha(R)$ versus $\ln (\frac{R}{R_m})$ at each time step is then calculated; hereafter we refer to this slope simply as $\delta_{\alpha,P}$. The evolution of $\delta_{\alpha,P}$ with respect to $\frac{t}{t_{rh}(t)}$ (which has been calculated using the cluster's projected half-mass radius) and the global mass function is plotted in Figure \ref{fig:dalphap}. It should be noted that to better compare with observations, we have used the more commonly observed mass range $0.3 < m < 0.8$ $M_{\odot}$ to illustrate the evolution of $\delta_{\alpha,P}$.

From Figure \ref{fig:dalphap} we see that the general evolution of $\delta_{\alpha,P}$ is nearly identical to $\delta_{\alpha}$ to within a scaling factor. The scaling factor is due to the fact that projection will lead to more low-mass stars \textit{appearing} to be located in the inner regions of the cluster thus minimizing the inner $\alpha(R)$ profile. Hence $\delta_{\alpha,P}$ decreases less rapidly than its three-dimensional counterpart with respect to both $\frac{t}{t_{rh}(t)}$ and $\alpha_G$.

In a future study we will carry out a dedicated investigation to compare the results of our simulations with observational data from a few wide field studies of Galactic globular clusters. We will further study all the aspects related to a direct comparison between observations and the results of numerical simulations. Factors such as the radial extent of observations, the range of masses observed, the completeness level of the observations can affect the values of $\delta_{\alpha,P}$  and their relation with the theoretical estimates of this quantity will all be considered.

\section{Discussion and Conclusions} \label{s_discussion}

Motivated by recent observational studies that have been able to measure the radial variation in the stellar mass function of globular clusters, we have studied the time evolution of the radial variation of the stellar mass function's slope, $\alpha(r)$, for a large suite of $N$-body star cluster simulations. In particular, we have focussed on how different initial sizes, masses, binary fractions, primordial mass segregation, black hole retention fractions, IMFs and external tidal fields affect the relationship between the time evolution of $\alpha(r)$ and the slope of the global mass function $\galph$ measured over different mass ranges. We find that:

\begin{enumerate}

\item In general the time evolution of $\alpha(r)$ is governed by the effects of internal two-body relaxation driving the system towards partial energy equipartition and mass segregation while the evolution with $\galph$ is a consequence of the preferential escape of low-mass stars associated with two-body relaxation-driven mass loss. Our simulations show that the evolution of model clusters in the $\dalph$-$\galph$ plane can be separated into three distinct stages (see Figure \ref{fig:schematic}): 

\begin{figure}
\centering
\includegraphics[width=\columnwidth]{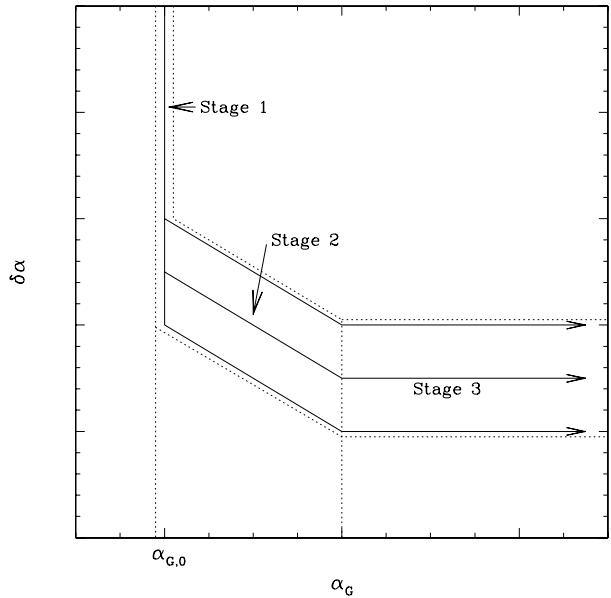}
\caption{An schematic diagram illustrating the evolution of $\dalph$ with respect to $\alpha_G$ for three clusters with different initial tidal filling factors. The evolution of $\dalph$ has been broken down into three stages that are described in Section \ref{s_discussion}.}
  \label{fig:schematic}
\end{figure}

\begin{itemize}

\item During the first stage, $\alpha_G$ does not evolve as the cluster is either not losing stars (because it is strongly under-filling) or is losing stars over the entire mass spectrum (because it is tidally filling but has not undergone mass segregation yet). $\dalph$ will continually decrease as the effects of two-body relaxation lead to mass segregation and, as a result, $\alpha(r)$ in the inner and outer radial bins increasing and decreasing respectively. The rate at which $\dalph$ decreases depends on the cluster's $t_{rh}$ and its subsequent evolution; clusters with short relaxation times will segregate and evolve towards smaller values of $\dalph$ more rapidly than clusters with long relaxation times.

\item The second stage begins when the joint effects of mass segregation and stellar escape due to two-body relaxation become significant. As the cluster continues its evolution and starts to lose preferentially low-mass stars as a result of the effects of two-body relaxation, $\alpha_G$ begins to increase and the external tidal field begins affecting $\alpha(r)$ in the outer regions. During this stage, $\dalph$ decreases at a slower rate because the rate at which low-mass stars segregate outward is balanced by the escape of preferentially low-mass stars from the cluster such that $\alpha(r)$ in the outermost regions stops decreasing.

\item During the third and final stage, the preferential escape of low-mass stars from the cluster becomes dominant over the segregation of low-mass stars outwards and $\alpha(r)$ in the outer regions begins increasing such that $\dalph$ remains constant in time and $\galph$ continues to increase. 

\end{itemize}

\item The minimum $\dalph$ reached by a cluster is primarily dependent on its $t_{rh}$ and how much it is able to segregate before stellar escapes causes $\alpha(r)$ in the outer regions to start increasing. The fact that all our model clusters reach similar minimum values of $\dalph$ may be due to clusters not actually being able to evolve to a state of pure energy equipartition \citep{merritt81, miocchi06, trenti13, gieles15, bianchini16}, with the minimum $\dalph$ representing the highest state of energy equipartition a star cluster can reach.

\item The position of a cluster on both the $\dalph$-$t_{rh}(t)$ and $\dalph$-$\galph$ planes at a given age can provide a qualitative indication of its degree of primordial mass segregation and initial relaxation time relative to its present day properties or to other clusters. Clusters that have undergone significantly more segregation than their current $t_{rh}(t)$ or $\galph$ would suggest likely had a shorter relaxation time in the past, were subject to a weaker tidal tidal field than their current orbit imposes, or were born primordially mass segregated.

\item Clusters that have different IMFs will be offset by their initial IMF slopes on the $\dalph$-$\galph$ plane such that the degeneracy between variations in the slope of the present-day mass functions of globular clusters that are due to dynamical evolution and those due to a non-universal IMF can be lifted. 

\item Clusters with non-zero black hole retention fractions will undergo less segregation, and therefore have less negative values of $\delta_\alpha$, than clusters which retain no black holes due to the black hole sub-population causing the cluster's core radius and tidal filling factor to increase.

\item When measured in projection, clusters appear less segregated and therefore have less negative values of $\delta_\alpha$ for a given $\frac{t}{t_{rh}(t)}$ and $\galph$. However, since the shape of the $\delta_\alpha$ evolutionary tracks remain unchanged in projection, observations can still be used to constrain a cluster's IMF and draw qualitative conclusions regarding its degree of primordial mass segregation, initial relaxation time and initial tidal filling factor relative to its present day orbit and structure.

\end{enumerate}

In light of the results shown by our theoretical study and the information that can be revealed by a full characterization of the radial variation 
of the slope of the PDMF, it is extremely important to further pursue wide field studies of Galactic globular clusters. In order to apply the results of our study to observations of globular clusters, special care must be taken to ensure models and observations are being compared over similar stellar mass ranges. However once these factors are taken into consideration, our study allows for the age, $\alpha(r)$, and $\alpha_G$ of a cluster to be used as a powerful tool for constraining the formation conditions and dynamical history of both globular clusters and their host galaxies. 

\section*{Acknowledgements}

This work was made possible in part by the facilities of the Shared Hierarchical Academic Research Computing Network (SHARCNET:www.sharcnet.ca) and Compute/Calcul Canada, in part by Lilly Endowment, Inc., through its support for the Indiana University Pervasive Technology Institute, and in part by the Indiana METACyt Initiative. The Indiana METACyt Initiative at IU is also supported in part by Lilly Endowment, Inc.


\bsp

\label{lastpage}

\end{document}